\documentclass[notitlepage,aps,amsmath,amssymb,graphicx,onecolumn]{revtex4-1}
\usepackage{dcolumn}% Align table columns on decimal point
\usepackage{bm}% bold math
\usepackage{float} %図用
\usepackage[dvipdfmx]{graphicx} %同上------------
\usepackage[top=30truemm,bottom=30truemm,left=25truemm,right=25truemm]{geometry}

\begin{document}
%\bibliographystyle{jplain}
%\preprint{APS/123-QED}

%\title{An Efficient Simulation Protocol for Determining
%\\the Density of States of Systems}% Force line breaks with \\
\title{An Efficient Simulation Protocol for Determining the Density of States:%}% Force line breaks with \\
\\Combination of Replica-Exchange Wang-Landau Method
\\and Multicanonical Replica-Exchange Method}

\author{Takuya Hayashi$^{1}$\thanks{tahayashi@tb.phys.nagoya-u.ac.jp} and 
%\author{Takuya Hayashi$^{1,}$\footnote{tahayashi@tb.phys.nagoya-u.ac.jp} and 
%%\altaffiliation{Department of Physics, Graduate of Science, Nagoya Universit, Nagoya, Aichi 464-8602, JAPAN. }
%%\email{tahayashi@tb.phys.nagoya-u.ac.jp}
Yuko Okamoto$^{1,2,3,4,5}$\thanks{okamoto@tb.phys.nagoya-u.ac.jp}}
%Yuko Okamoto$^{1,2,3,4,5,}$\footnote{okamoto@tb.phys.nagoya-u.ac.jp}}
%\altaffiliation{Department of Physics, Graduate of Science, Nagoya Universit, Nagoya, Aichi 464-8602, JAPAN: 
%Structural Biology Research Center, Graduate School of Science, Nagoya University, Nagoya, Aichi 464-8602, Japan:
%Center for Computational Science, Graduate School of Engineering, Nagoya University, Nagoya, Aichi 464-8603, Japan: 
%Information Technology Center, Nagoya University, Nagoya, Aichi 464-8601, Japan, JST-CREST, Nagoya, Aichi 464-8602, Japan}
%\email{okamoto@tb.phys.nagoya-u.ac.jp}
%\inst{
\affiliation{
$^{1}$Department of Physics, Graduate School of Science, Nagoya University, Nagoya, Aichi 464-8602, Japan \\
$^{2}$Structural Biology Research Center, Graduate School of Science, Nagoya University, Nagoya, Aichi 464-8602, Japan \\
$^{3}$Center for Computational Science, Graduate School of Engineering, Nagoya University, Nagoya, Aichi 464-8603, Japan \\
$^{4}$Information Technology Center, Nagoya University, Nagoya, Aichi 464-8601, Japan \\
$^{5}$JST-CREST, Nagoya, Aichi 464-8602, Japan
}



%\vskip 1.5 cm
%\vfill
%~\\
%~\\

\begin{abstract}
By combining two generalized-ensemble algorithms, Replica-Exchange Wang-Landau (REWL)
method and Multicanonical Replica-Exchange Method (MUCAREM), 
we propose an effective simulation protocol
%to determine the density of states of a system.
%to determine the density of states of complex systems with high accuracy.
to determine the density of states with high accuracy.
The new protocol is referred to as REWL-MUCAREM, and REWL is first performed and then
MUCAREM is performed next.
In order to verify the effectiveness of our protocol, 
we performed simulations of a square-lattice Ising model by the three methods,
namely, REWL, MUCAREM, and REWL-MUCAREM. 
The results showed that the density of states obtained by the 
REWL-MUCAREM is
more accurate than that is estimated by the two methods separately. 
\end{abstract}

\pacs{Valid PACS appear here}% PACS, the Physics and Astronomy

\maketitle
%========================================
\section{Introduction}
%\hspace{8mm}
The statistical mechanical expectation value of a physical quantity 
can be accurately calculated if the density of states (DOS) is given.
However, in many cases, we do not know DOS \it{a priori} \rm
and it is often difficult to obtain it theoretically or experimentally.
In recent decades, many methods were developed for the determination of DOS 
by using Monte Carlo (MC) and/or molecular dynamics (MD) simulations
(e.g., see Refs.~\cite{UMBRELLA}--\cite{R12c}).
One of the earliest such methods may be {\it Umbrella Sampling} \cite{UMBRELLA}.
{\it Muticanonical Algorithm} \cite{R1}--\cite{R3}, {\it Simulated Tempering}
\cite{ST1}--\cite{ST3}, {\it Replica-Exchage Method} \cite{R6}--\cite{R7},
{\it Wang-Landau} method \cite{R4a,R4b}, and {\it Metadynamics} 
\cite{META}--\cite{META3} were then developed, and
generalizations and extensions of these methods were further proposed
\cite{R7b}--\cite{R12c}.
These methods are closely related. For example, 
it has been shown that 
{\it Statistical Temperature Molecular Dyanamics} \cite{STMD}
is equivalent to Metadynamics \cite{RESTMD_META}. We also remark that 
Metadynamics can be considered to be Wang-Landau method 
in reaction coordinate space (rather than energy space) \cite{RevYO}.
%\cite{R1,R2,R3,R4a,R4b,R6,R6a,R6b,R7,R7b,R8,R9,R10,R10b,R11,R12,R12c} 
%and there are countless algorithms associating with these methods 
%(e.g.\cite{UMBRELLA,META,STMD,TWL,RESTMD,RESTMD_META}).
These methods have been successfully applied to a wide range of problems in condensed matter and statistical physics including spin glasses, liquid crystals, polymers, and proteins.
Nevertheless, the problem still remains that the exact calculation of DOS cannot be 
achieved when the systems become large and complex. 

In this article, we propose an efficient simulation protocol to obtain 
the most precise DOS 
by combining the {\it Replica-Exchange Wang-Landau} method (REWL)\cite{R11,R12}
%\sl Replica-Exchange Wang-Landau \rm method (REWL)\cite{R11,R12,R13}
and the {\it Multicanonical Replica-Exchange Method} (MUCAREM)\cite{R8,R9,R10}. 
  
This article is organized as follows. 
In Sec.~II, we explain the methods. 
In Sec.~III, the computational details are given. 
In Sec.~IV, the results and discussion are presented.
Sec.~V is devoted to conclusions. \\
%========================================
\section{Computational methods}
We first introduce three basic generalized-ensemble algorithms, {\it Multicanonical Algorithm}, {\it Wang-Landau} method, and {\it Replica-Exchange Method}.
%MUCA
The {\it Multicanonical Algorithm} (MUCA)\cite{R1,R2} is one of the representative methods. 
A simulation in multicanonical ensemble is based on a non-Boltzmann 
weight factor, which we refer to as the multicanonical weight factor.
This is inversely proportional to DOS of the system, 
and a free random walk in potential energy space is realized 
so that a wide configurational space may be sampled.
The DOS is often not known {\it a priori}. 
The multicanonical weight factor is usually determined by iterations of short trial simulations\cite{R3,R20}. 
After a production run with the determined MUCA weight factor, the single-histogram reweighting techniques\cite{R14a} are employed to obtain an accurate DOS.
However, the weight factor determination process can be tedious and difficult.
%WL
The {\it Wang-Landau} method (WL)\cite{R4a,R4b} solved this problem drastically. 
In the WL sampling, the weight factor, which is also inversely proportional to DOS, is updated during the simulation by multiplying a constant to the weight factor. 
This procedure leads to a uniform histogram in potential energy space, 
and the modified weight factor converges to the inverse of the DOS.
%REM
Another powerful algorithm is the {\it Replica-Exchange Method} (REM)\cite{R6,R6a} 
(it is also referred to as parallel tempering\cite{MPRL}). 
Closely related method was independently developed in~\cite{SW}.
In this method, several copies (replicas) of the original system at different temperatures are
simulated independently and simultaneously by conventional canonical MC or MD.
Every few steps, pairs of replicas are exchanged with a specified transition probability.
This exchange process realizes a random walk in temperature space, which in turn
induces a random walk in potential energy space.
After a production simulation, 
the multiple histogram reweighting techniques\cite{R14b,R15} (an extension of which is also 
referred to as the Weighted Histogram Analysis Method (WHAM)\cite{R15}) are used 
in order to determine the most accurate DOS from all the histograms of sampled potential 
energy at different temperatures.

%========================================
%\hspace{8mm}
These basic simulation methods can be combined for more effective sampling. 
One method is referred to as the {\it Multicanonical Replica-Exchange Method} 
(MUCAREM)\cite{R8,R9,R10}.
In this method, the total energy range where we want to calculate the DOS is divided into smaller 
regions, each corresponding to a replica, and MUCA simulations are performed 
independently and simultaneously in each replica.
Every few steps, a pair of neighboring replicas are exchanged like REM.
The configurations can be sampled more effectively than ordinary MUCA because of replica exchange.
The final, most accurate estimation of DOS is obtained by the multiple-histogram reweighting techniques again~\cite{R8,R9,R10}. 
A similar method is the {\it Replica-Exchange Wang-Landau} (REWL) method\cite{R11,R12}.
The idea is almost the same as in MUCAREM except for using WL instead of MUCA for each replica.
After a REWL simulation, DOS pieces are obtained for different energy regions.
Connecting these pieces at the point where the slope of DOS is coincident, 
we can obtain the final estimation of DOS over the entire energy range. 

%\hspace{8mm} 
We found that the DOS with the highest accuracy can be obtained by combining these two methods. 
The REWL is employed in the first half of the total number of MC (or MD) steps 
in order to get a rough estimate of MUCAREM weight factor 
and the MUCAREM is performed in the second half in order to refine the DOS. 
We refer to this protocol as REWL-MUCAREM. 
The DOS thus obtained has higher accuracy than that is estimated by the 
two methods separately. 

%========================================
%\hspace{8mm}
A brief explanation of MUCA is now given here. 
The multicanonical probability distribution of potential energy $P_{{\rm MUCA}}(E)$ is defined by
\begin{equation}
 \label{EQ1}
%\begin{eqnarray}
 P_{{\rm MUCA}}(E) \propto  g(E) W_{{\rm MUCA}}(E) \equiv {\rm const}~,
%\end{eqnarray}
\end{equation}
where $W_{{\rm MUCA}}(E)$ is the multicanonical weight factor and
the function $g(E)$ is the DOS. $E$ is the total potential energy of a system.
By omitting a constant factor, we have 
\begin{eqnarray}
 \label{EQ2}
 \displaystyle
 W_{{\rm MUCA}}(E) &= \displaystyle{\frac{1}{g(E)}}~.
\end{eqnarray}
In MUCA MC simulations, the trial moves are accepted with the following Metropolis transition probability
 $w \left( E \rightarrow E' \right)$:
\begin{eqnarray}
 \label{EQ3}
w \left( E \rightarrow E' \right) = {\rm min} \sl \left[ 1, \displaystyle \frac{W_{{\rm MUCA}} \sl (E')}
                                                                             {W_{{\rm MUCA}} \sl (E)} \right] 
                                            = {\rm min} \sl \left[ 1, \displaystyle \frac{g(E)}{g(E')} \right]~.
\end{eqnarray}
Here, $E$ is the potential energy of the original configuration and $E'$ is that of a proposed one.
After a long production run, the best estimate of DOS can be obtained by the single-histogram reweighting techniques:
\begin{eqnarray}
 \label{EQ4}
g(E)= \displaystyle \frac{H(E)}{W_{{\rm MUCA} }\sl (E)}~,
\end{eqnarray}
where $H(E)$ is the histogram of sampled potential energy. 
Practically, the $W_{{\rm MUCA}} (E)$ is set $\exp [-\beta E ]$ at first and 
modified by repeating sampling and reweighting. 
Here, $\beta$ is the inverse of temperature $T$ ($\beta = 1/ k_{{\rm B}} T$ with $k_{{\rm B}}$ being the Boltzmann constant).

%\hspace{8mm}
The WL also uses $1/g(E)$ as the weight factor and the Metropolis criterion is the same as in Eq.(\ref{EQ3}). 
However, $g(E)$ is updated dynamically as $g(E) \rightarrow f \times g(E)$ during the simulation 
when the simulation visits a certain energy value $E$. 
$f$ is a modification factor. 
We continue the updating until the energy histogram becomes flat. 
If $H(E)$ is flat enough, a next simulation begins after resetting 
the histogram to zero and reducing the modification factor (usually, $f \rightarrow \sqrt{f}$). 
The flatness evaluation can be done in various ways. 
%In this article we considered that the histogram is sufficiently flat when the criteria 
%$H_{\rm min}/H_{\rm max} > 0.5~$ was satisfied,
In this article, we considered that the histogram is sufficiently flat when
\begin{eqnarray}
 \label{EQ5}
 \displaystyle
 \frac{H_{\rm min}}{H_{\rm max}} > 0.5~,
\end{eqnarray}
where $H_{\rm min}$ and
$H_{\rm max}$ are the least number and the largest number of nonzero entries in the histogram,
respectively\cite{R16}.
This process is terminated when the modification factor attains a predetermined 
value $f_{{\rm final}}$ and  $\exp(10^{-8}) \simeq 1.000\,000\,01$ is often used as $ f_{{\rm final}}$.
Hence, the estimated $g(E)$ tends to converge to the true DOS of the system
within this much accuracy set by $f_{{\rm final}}$. 
(Several reports argue that there is a possibility that conventional WL 
algorithm has a systematic error which does not decrease any more. See,
e.g., Ref.\cite{TWL}.)

%========================================
%\hspace{8mm}
In MUCAREM, the entire energy range of interest $\left[ E_{{\rm min}},E_{{\rm max}} \right]$ 
is divided into $M$ sub-regions, $ E_{{\rm min}}^{\{m\}} \leq E \leq E_{{\rm max}}^{\{m\}}$ 
$(m=1,2,\cdots,M)$, where  
$E_{{\rm min}}^{\{1\}} = E_{\rm min}$ and $E_{{\rm max}}^{\{M\}}=E_{\rm max}$. 
There should be some overlaps between the adjacent regions.
MUCAREM uses $M$ replicas of the original system. 
The weight factor for sub-region $m$ is defined by\cite{R8,R9,R10}: 
\begin{eqnarray}
 \label{EQ6}
    W^{\{m\}}_{{\rm MUCA}} (E) = 
  \displaystyle{
   \begin{cases}
     e^{- \beta_{\rm L}^{\{m\}} E},      & {\rm for} \, \, E < E^{\{m\}}_{{\rm min}} \,\,\,\, ,  \\ 
     \displaystyle{\frac{1}{g_m(E)}},   & {\rm for} \, \, E^{\{m\}}_{{\rm min}} \le E \le E^{\{m\}}_{{\rm max}} 
\,\,\,\, ,  \\
     e^{- \beta_{\rm H}^{\{m\}} E},      & {\rm for} \, \, E > E^{\{m\}}_{{\rm max}} \,\,\,\, ,  
   \end{cases}
  }
\end{eqnarray}
where $g_m(E)$ is the DOS for 
$E^{\{m\}}_{{\rm min}} \le E \le E^{\{m\}}_{{\rm max}}$ 
in sub-region $m$,
$\beta_{\rm L}^{\{m\}} = d \log \left[g_m(E) \right] / d E ~(E=E^{\{m\}}_{\rm min})$
and, $\beta_{\rm H}^{\{m\}} = d \log \left[ \it g_m(E) \right] / d E ~(E=E^{\{m\}}_{\rm max})$.
The MUCAREM weight factor $W_{{\rm MUCAREM}}(E)$ for the entire energy range is expressed by the following formula:
\begin{eqnarray}
 \label{EQ7}
 W_{{\rm MUCAREM}}(E) =     & \displaystyle{ \prod_{m=1}^{M} } W_{{\rm MUCA}}^{\{m\}}(E)~.
\end{eqnarray}
After a certain number of independent MC steps, replica exchange is proposed 
between two replicas, $i$ and $j$, in neighboring sub-regions, $m$ and $m+1$, respectively. 
%The energy of a replica $m$ is $E_{m}$ and the temporally DOS is $g_{m}(E)$. 
The transition probability, $w_{{\rm MUCAREM}}$, of this replica exchange is given by
\begin{eqnarray}
 \label{EQ8}
  w_{{\rm MUCAREM}} &=& \displaystyle 
        {\rm min} \sl \left[ 1 , \frac{W^{\{m\}}_{\rm MUCA}(E_j) W^{\{m+1\}}_{\rm MUCA}(E_i)}  
                                      {W^{\{m\}}_{\rm MUCA}(E_i) W^{\{m+1\}}_{\rm MUCA}(E_j)}  \right] ,
\end{eqnarray}
where $E_i$ and $E_j$ are the energy of replicas $i$ and $j$ before the replica exchange,
respectively.
If replica exchange is accepted, the two replicas exchange their weight factors 
$W^{\{m\}}_{{\rm MUCA}} (E)$ and  
$W^{\{m+1\}}_{{\rm MUCA}} (E)$   
and energy histogram $H_{m}(E)$ and $H_{m+1}(E)$. 
The final estimation of DOS can be obtained from $H_{m}(E)$ after a simulation 
by the multiple-histogram reweighting techniques or WHAM.
Let $n_{m}$ be the total number of samples for the $m$-th energy sub-region.
The final estimation of DOS, $g(E)$, is obtained by solving the following WHAM equations 
self-consistently by iteration\cite{R9}:
\begin{eqnarray}
 \label{EQ9}
   \left\{
   \begin{array}{l}
      g(E) = \frac{ \displaystyle{ \sum_{m=1}^{M} } H_{m} (E) }
              {\displaystyle{ \sum_{m=1}^{M}} n_{m} \exp \left( f_{m} \right) W_{{\rm MUCA}}^{\{m\}} (E)} ~, \\ \\
      \exp \left( - f_{m} \right) =\displaystyle{ \sum_{E} g(E) W_{{\rm MUCA}}^{\{m\}}(E) }~.
   \end{array}
   \right.
\end{eqnarray}
These MUCAREM sampling and WHAM reweighting processes can, in principle, be repeated to obtain more accurate DOS\cite{R10}. 
We remark that REM is often used to obtain the first estimate of DOS in the MUCAREM iterations.
We also remark that when REM instead of MUCAREM is performed, the best
estimate of DOS can be obtained by solving Eq.~(\ref{EQ9}), where 
$W^{\{m\}}_{{\rm MUCA}} (E)$ is replaced by 
$\exp ( - \beta_{m} E)$ with temperature $T_m~(\beta_{m} = 1/k_{\rm B}T_{m})$
for ($m=1, 2, \cdots M$).
%========================================
%\begin{table*}[hpbt]
\begin{table*}[t]
\caption{\label{tab:table1}Conditions of the present simulations.
%\textbackslash
}
%\begin{ruledtabular}
\begin{tabular}{ccccc}
 \hline 
 \hline 
%    \shortstack{Method}
%  &\shortstack{number of replicas}
%  &\shortstack{frequency of \\ Wang - Landau criterion}
%  &\shortstack{the number of \\ total MC sweeps}\\
      
   & 
   & 
   & Frequency (in MC sweeps) \rule[0mm]{0mm}{4mm}
   &                              \\
      Methods
   & Number of spins\,\,\,
   & Number of replicas\,\,\, 
   & of flatness evaluation\,\,\,
   & Total MC sweeps     \\
   & $N$ 
   & $M$ 
   & in Eq.~(\ref{EQ5})
   & per replica                             \\
 \hline 
%\shortstack{REWL}
%  &$8,16,32,64$
%  &$200$ 
%  &$200,000$ \\ \\
%\shortstack{MUCAREM}
%  &$8,16,32,64$
%  & NA
%  &$200,000$ \\ \\
%\shortstack{REWL-MUCAREM}
%  &$\shortstack{8,16,32,64 - 8,16,32,64}$
%  &$\shortstack{200 - NA}$ 
%  &$\shortstack{100,000 - 100,000}$ \\
MUCAREM & 64              & 4        & NA   & 200\,000\rule[0mm]{0mm}{4mm}\\
         & 256            &  8       &      & 200\,000 \\
         & 1024           & 16       &     & 200\,000 \\
         & 4096           & 32       &       & 500\,000 \\
         & 16384         & 64       &      & 3\,000\,000 \\
REWL & 64       & 4         & 1000  & 200\,000 \\
                &  256           &  8       &      & 200\,000 \\
                & 1024           & 16       &      & 200\,000 \\
                & 4096           & 32       &    & 500\,000 \\
                & 16384         & 64       &     & 3\,000\,000 \\
REWL$-$MUCAREM & 64              &  4        & 1000$-$NA  & 100\,000$-$100\,000 \\
                              &  256           &  8      &                 & 100\,000$-$100\,000 \\
                              & 1024           & 16      &                 & 100\,000$-$100\,000 \\
                              & 4096           & 32      &                & 250\,000$-$250\,000 \\
                              & 16384         & 64      &                & 1\,500\,000$-$1\,500\,000 \\
 \hline 
 \hline 
\end{tabular}
%\end{ruledtabular}
\end{table*}
%========================================

The REWL method is essentially based on the same weight factors as in MUCAREM, while the WL simulations replace the MUCA simulations for each replica.
This simulation is terminated when the modification factors on all sub-regions attain a 
certain minimum value $f_{{\rm final}}$. 
After a REWL simulation, $M$ pieces of DOS fragments with overlapping energy intervals are obtained. 
The fragments need to be connected in order to determine 
the final DOS in the entire energy range $\left[ E_{{\rm min}} ,E_{ {\rm max}} \right]$. 
The joining point for any two overlapping DOS pieces is chosen 
where the inverse microcanonical temperature 
$\beta = d \log \left[ \it g(E) \right] / d \it E$ coincides best\cite{R11,R12}.

%========================================
\section{Computational details}
%\hspace{8mm}
In order to compare the effectiveness of the REWL-MUCAREM with other methods, 
we performed simulations of a 2-dimensional Ising model with periodic boundary conditions.

In a square-lattice Ising model, the total energy $E$ is defined by 
\begin{eqnarray}
 \label{EQ10}
\displaystyle
E=-J \sum_{ \langle i,j \rangle} S_{i} S_{j}~,
\end{eqnarray} 
where $i$ and $j$ are labels for lattice points.
$J$ is the magnitude of interaction between neighboring spins.
In this article, $J$ and $k_{{\rm B}}$ are set to one for simplicity.
$\langle i,j \rangle$ represents pairs of nearest-neighbor spins.
$S_{i}$ is the state of spin on a lattice point $i$ and takes on values of $\pm 1$.
Beale calculated the exact DOS of the model of finite sizes\cite{R17,R18}.

%\hspace{8.0mm}
Table I lists the conditions of our simulations.
The total number of spins $N$ is $L^2$, where $L$ is the length of a side of the square lattice,
The total number of spins considered was $N=64, 256, 1024, 4096$, and $16384$.
One MC sweep is defined as an evaluation of Metropolis criteria $N$ times.
The cost of computations (for example, the total number of MC sweeps) was set equal.
However, we should point out that while the ordinary REWL algorithm is terminated when the recursion 
factor $f$ converged to $f_{\rm final}$, but our REWL simulations were finished after a certain 
fixed number of flatness evaluations had been made.

A Marsaglia random number generator was employed 
and we used the program code on open source\cite{R19,R20}.
The number of replicas was set equal to $L/2$.
Each replica performed a MUCA simulation in MUCAREM or a WL simulation in REWL 
within their energy sub-regions, which had an overlap of about $80$ percent between neighboring sub-regions.
%In the cases of REWL and REWL-MUCAREM simulations, the WL flatness criterion in Eq.~(\ref{EQ5}) was tested every $1000$ MC sweeps.
In the cases of REWL and REWL-MUCAREM simulations, WL flatness criterion was tested every $1000$ MC sweeps.
If the histogram of energy distribution is sufficiently flat at this time, the WL recursion factor was reduced. 
Replica exchange was tried every $100$ MC sweeps.
%This means that replica exchange was tried $2000$ times altogether for the systems up to $N=1024$, 
%$5000$ times for $N=4096$, and $30000$ times for $N=16384$ (see Table I).
The cost of calculation in our simulations was measured by the total number of MC sweeps 
because we spend most of computational time to perform MC simulations.
With the conditions in Table I, we made $n=25$ independent runs with different 
initial random number seeds. %in order to estimate errors (and we obtained DOS estimates $25$ times for each set of conditions).
In this work, we did not iterate the DOS evaluation during the MUCAREM 
simulations for simplicity.
In the present MUCAREM simulation, the first half of the total MC sweeps
was run with REM and the remaining of the simulation was MUCAREM with the DOS obtained from the REM simulation.
We evaluated the effectiveness of iterations of MUCAREM and WHAM (see Appendix A).
In the REM simulation, $M$ temperature values were evenly distributed
between $\beta_1 = 1.0$ and $\beta_M = 0.01$.
%Finally, in REWL-MUCAREM, MUCAREM simulation was performed with the DOS determined by REWL.
%*****************************
%\begin{figure}[htbp]
%  \begin{center}
%    \begin{tabular}{cc}
% \begin{minipage}{0.45\hsize}
%  \begin{center}
%  \mbox{\raisebox{-40.mm}{\includegraphics[width=70mm,height=70mm,clip]{SPHEAT.eps}}}
%  \end{center}
%  \caption{The specific heat. (a) is the exact solutions which were calculated by exact DOS\cite{R17,R18}. (b), (c), and (d) were obtained by simulations with $L=8,$ 16, 32, 64, and 128 
%by MUCAREM, REWL, and REWL-MUCAREM, respectively.}
%  \label{fig1_a}
% \end{minipage}
% \begin{minipage}{0.45\hsize}
%  \begin{center}
%   \includegraphics[width=70mm,height=60mm,clip]{DIFF_SPC.eps}
%  \end{center}
%  \caption{The differences of specific heat between the simulation results and exact one in Eq~(\ref{EQ15}).}
%  \label{fig1_b}
% \end{minipage}
%    \end{tabular}
%\caption{The specific heat. (a) is the exact solutions which were calculated by exact DOS\cite{R17,R18}. %(b), (c), and (d) were obtained by simulations with $L=8,$ 16, 32, 64, and 128 by MUCAREM, REWL, and %REWL-MUCAREM, respectively. Right figure shows that the differences between the simulation results %and exact one in Eq~(\ref{EQ15}).}
%  \end{center}
%\end{figure}

%\begin{figure}[hbt]
\begin{figure}[t]
%\begin{tabular}{p{0.4\textwidth}p{1.\textwidth}p{0.4\textwidth}}
\begin{tabular}{cc}
\centering\includegraphics*[width=80mm,height=85mm,keepaspectratio,clip]{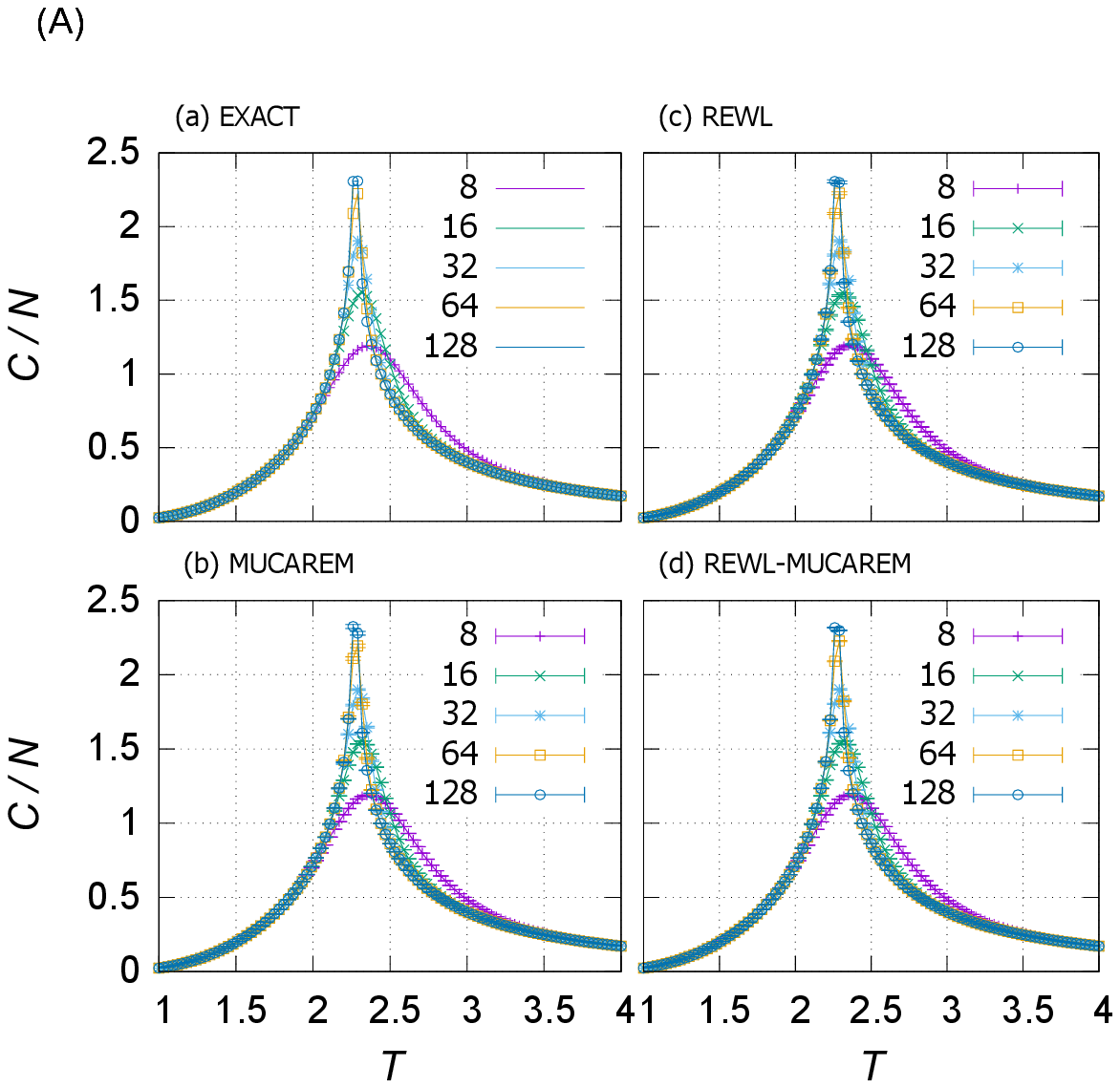}
%&&
\centering\includegraphics*[width=80mm,height=74mm,keepaspectratio,clip]{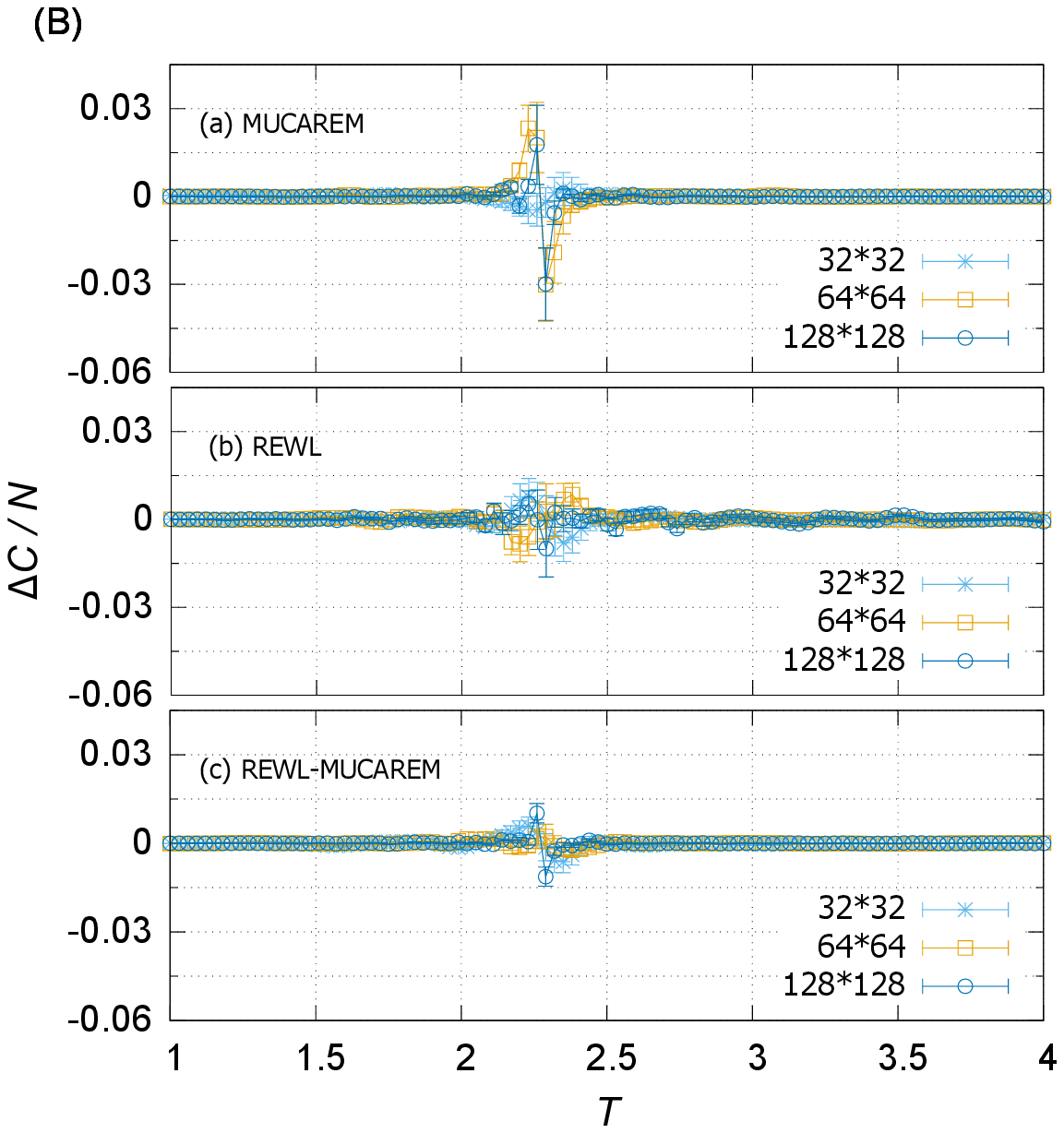}
\end{tabular}
 \caption{The specific heat. (a) in (A) gives the exact solutions 
which were calculated by the exact DOS\cite{R17,R18}. (b), (c), and (d) 
were obtained by simulations with $L=8, 16, 32, 64,$ and $128$ by MUCAREM, REWL, and REWL-MUCAREM, respectively. (B) shows the differences between the simulation results and exact one, $\Delta C(T) = C_{\rm sim}(T) - C_{\rm EXACT}(T)$.}
\end{figure}
%*****************************
%***************************** SPC_HEAT
\section{Results and discussion}
\hspace{8.0mm}
The four figures in Figs.~1(A) show the specific heat which was calculated from the estimated DOS by using the following equation:
\begin{eqnarray}
 \label{EQ11}
  C(T)=\displaystyle{ \frac{ \langle E^2 \rangle_T  - \langle E \rangle_T ^ 2}{T^{2}} }~,
\end{eqnarray}
where
\begin{eqnarray}
 \label{EQ12}
  \langle A \rangle_T =  \frac{ \displaystyle{ \sum_{E} } A(E) g(E) e^{- \beta E }}{\displaystyle \sum_{E} g(E) e^{- \beta E }}~,
\end{eqnarray}
and $A(E)$ is any physical quantity that depends on $E$.
%The average value and the errors were obtained by the standard error estimation:
The errors were obtained by the standard error estimation:
\begin{equation}
 \label{EQ14}
  \varepsilon_A = \sqrt{\frac{
\displaystyle \sum_{i=1}^{n}  \left( A^{\left\{ i \right\}} - 
\overline{A} \right)^{2} }{n \left( n-1 \right)}} \rule[0mm]{0mm}{8mm}~~~~~,~~~~~
\overline{A} = \frac{ \displaystyle{\sum_{i=1}^{n}} A^{\left\{ i \right\}} }{n}~.
\end{equation}
Here, $A^{\left\{ i \right\}}$ is obtained from the $i$-th simulation 
$(i=1,2,\cdots,n)$.
%Fig.~$1$ shows the specific heat which was calculated from the estimated DOS by using the following equation:
%\begin{eqnarray}
% \label{EQ11}
%  C(T)=\displaystyle{ \frac{ \langle E^2 \rangle_T  - \langle E \rangle_T ^ 2}{T^{2}} }~,
%\end{eqnarray}
%where
%\begin{eqnarray}
% \label{EQ12}
%  \langle A \rangle_T =  \frac{ \displaystyle{ \sum_{E} } A(E) g(E) e^{- \beta E }}{\displaystyle \sum_{E} g(E) e^{- \beta E }}~,
%\end{eqnarray}
%and $A(E)$ is any physical quantity that depends on $E$.
%The specific heat in Figs.~1(b)--1(d)  obtained from the simulations is defined by
%\begin{equation}
% \label{EQ13}
%  C_{\rm sim}(T) = \displaystyle \frac{1}{n} 
%\sum_{i=1}^{n} C_{\rm sim}^{\left\{ i \right\}}(T)~, 
%\end{equation}
%where $C_{\rm sim}^{\{ i \}}(T)$ is the specific heat calculated from Eqs.~$(11)$ and $(12)$
%for the $i$-th simulation ($i=1, 2, \cdots, n$). 
%The errors in Fig.~$1$ (and Fig.~2 below)  were estimated by the standard error:
%\begin{equation}
% \label{EQ14}
%  \varepsilon_C(T) = \sqrt{\frac{
%\displaystyle \sum_{i=1}^{n}  \left( C_{\rm sim}^{\left\{ i \right\}}(T) - 
%C_{\rm sim}(T) \right)^{2} }{n \left( n-1 \right)}}~. \rule[0mm]{0mm}{8mm}
%\end{equation}
%
Although the exact values of specific heat in finite sizes were obtained by 
Ferdinand and Fisher\cite{R21}, 
we calculated the exact specific heat in Fig.~1(A)(a) from the exact DOS, 
$g_{{\rm EXACT}}(E)$, of Beale\cite{R17,R18} by using Eqs.~$(\ref{EQ11})$ and $(\ref{EQ12})$.
We used the Mathematica code, which is given in~\cite{R18}, for the calculations of $g_{{\rm EXACT}}(E)$.
All the algorithms could reproduce the exact solutions very well.

The differences between exact values and simulation results are shown in Fig.~1(B).
%It was calculated by the following equation:
%\begin{eqnarray}
% \label{EQ15}
%  \Delta C(T) = C_{\rm EXACT}(T) - C_{\rm sim}(T)~.
%\end{eqnarray}
Note that $|\Delta C(T)|$ takes maximum values around the phase transition temperature $T_{c} = 2/\log (1 + \sqrt{2})\simeq 2.269$ in each method.
The results imply that the results of the three methods agree with the exact 
ones in the order of REWL-MUCAREM, REWL, and MUCAREM.
It means that REWL-MUCAREM could get more accurate DOS than the other two methods.

%========================================
\begin{figure}[t]
 \label{DOSZ}
\begin{centering}
 \includegraphics[width=12.0cm]{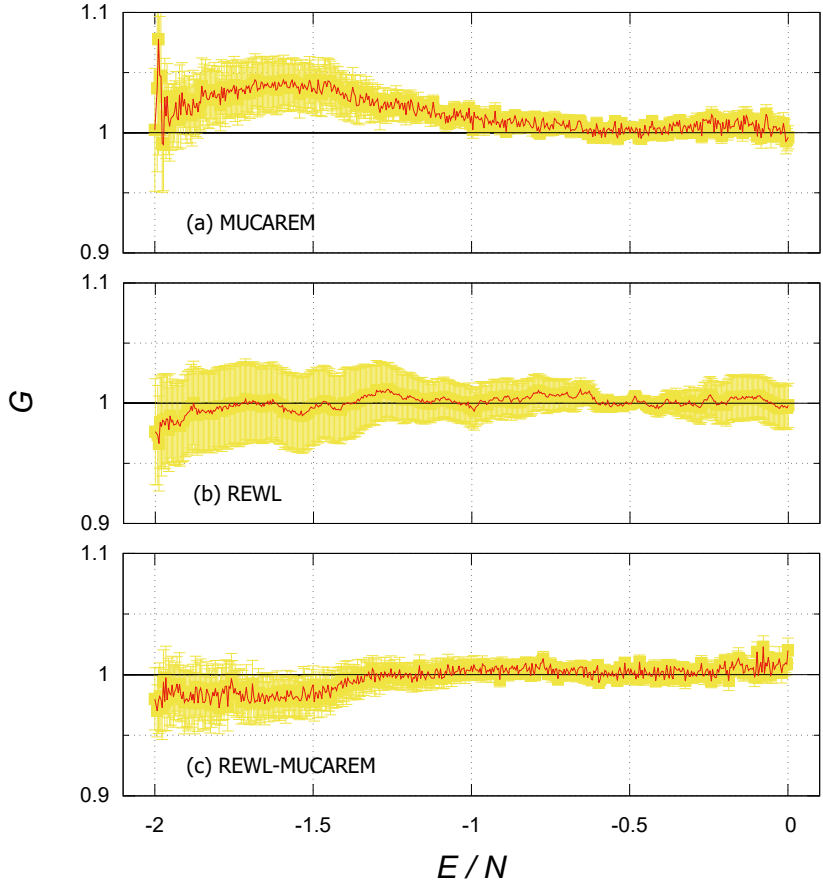}
  \caption{Mean local flatness $G(E)$ in Eq.~($16$) (red curves). The total number of spins is $N=32 \times 32$.
$G(E)$ were obtained from the simulations by (a) MUCAREM, (b) REWL, and (c) REWL-MUCAREM. 
The error bars (vertical yellow bars) were obtained by standard error estimation. The best estimated DOS will give $G(E) = 1$.}
\end{centering}
\end{figure}
%========================================
%*****************************
%*****************************DOS
%\hspace{8.0mm}
In order to directly compare the accuracy of DOS among the three methods, 
we show the mean local flatness $G(E)$ in Fig.~2 for the system of $N=32 \times 32$, where
\begin{eqnarray}
 \label{EQ16}
 \begin{cases}
  G(E) &= \displaystyle \frac{1}{n} \sum_{i=1}^{n} G^{\left\{ i \right\}}(E)~, \\
 \rule[0mm]{0mm}{8mm}
  G^{\left\{ i \right\}}(E) &= \displaystyle \cfrac{g_{{\rm sim}}^{\left\{ i \right\}}(E)}{g_{{\rm EXACT}}(E)}~.

 \end{cases}
\end{eqnarray}
Here, $g_{{\rm sim}}^{\left\{ i \right\}}$ is the DOS estimated from the $i$-th simulation $(i=1,2,\cdots,n)$. 
%In Ising model, there are two states at the ground state energy $E_{0}$, 
%which are all spins up or all spins down. Hence, $\log g_{{\rm EXACT}} (E_{0})$ 
%takes the value $\log 2$.
%We matched $\log g_{{\rm EXACT}}(E)$ and $\log g_{{\rm sim}}^{\left\{ i \right\}}(E)$ at $E/N=-0.25$.
We matched $\log g_{{\rm EXACT}}(E)$ and $\log g_{{\rm sim}}^{\left\{ i \right\}}(E)$ at $E/N=-0.5$.
If $g_{{\rm sim}}^{\left\{ i \right\}}(E)$ is equal to $g_{{\rm EXACT}}(E)$, $G^{\left\{ i \right\}}(E)$ 
becomes flat ($=1$) ideally in the entire energy range.
The red curves and the yellow vertical bars in Fig.~2 are the values of $G(E)$ and the error bars, respectively.
The errors were also estimated by the standard error estimation.
%The errors were also estimated by the standard error estimation in Eq. (\ref{EQ14}).
%\begin{equation}
% \label{EQ17}
%  \varepsilon(E) = \sqrt{\frac{
%\displaystyle \sum_{i=1}^{n}  \left( G^{\left\{ i \right\}}(E) - G(E) \right)^{2} }{n \left( n-1 \right)}}~. \rule[0mm]{0mm}{8mm}
%\end{equation}
The difference between red curve and black base line became large at lower energy region. 
The tendency became stronger in larger systems, where the phase transition became stronger (see Fig 1).
Because the error bars are the smallest at lower energy region among the three methods, 
REWL-MUCAREM could obtain more precise DOS than REWL and MUCAREM.

%========================================
\begin{figure}
  \centering
 \includegraphics[width=12.0cm]{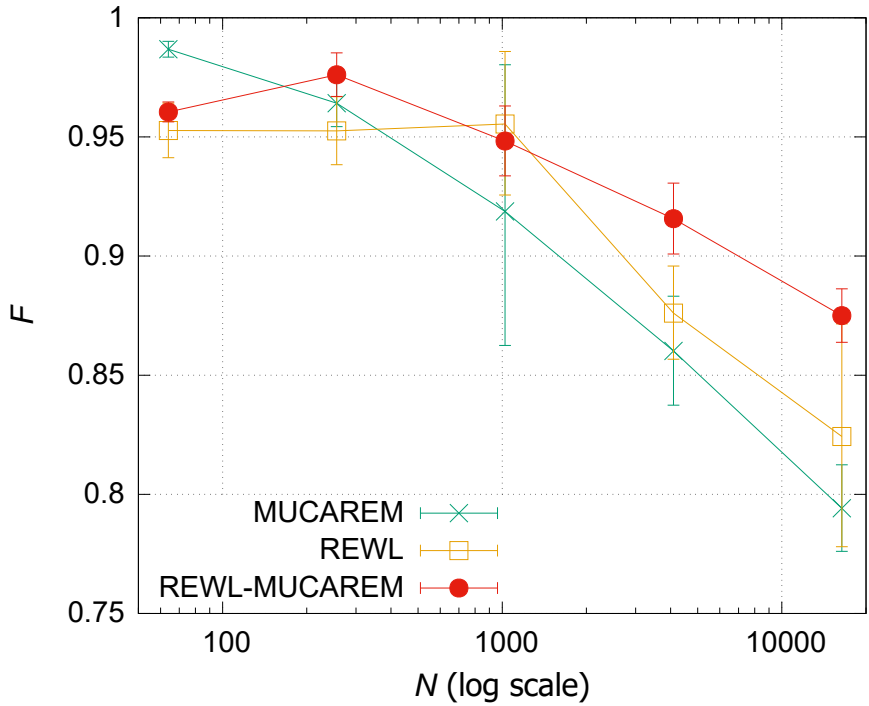}
  \caption{Global flatness $F$ defined in Eq.~(\ref{EQ18}) as a function of the total number of spins, $N$.
If $g_{{\rm sim}}^{\left\{ i \right\}}(E)$ is equal to $g_{{\rm EXACT}}(E)$ in the entire energy range, $F$ takes a value one.} 
%According to B.A. Berg, $F$ is enough to estimate the expectation values if it takes 0.5.}
\end{figure}
%========================================
%\hspace{8mm}
In order to examine the accuracy of DOS further, we define the degree of global flatness $F$ 
by the following formula:
\begin{eqnarray}
 \label{EQ18}
\displaystyle
F \equiv \frac{G_{\rm min}}{G_{\rm max}}~,
\end{eqnarray}
where $G_{\rm min}$ is the minimum value of $G(E)$ over the entire energy range and $G_{\rm max}$ is the maximum one.
$F$ takes on values between 0 and 1.
The closer the calculated $g_{\rm sim}^{\left\{ i \right\}} (E)$ is to $g_{\rm EXACT}(E)$ globally, the closer $F$ is to $1$.
Fig.~$3$ shows the measured flatness $F$.
As a measure of errors, we define the minimum value of $F$ and 
maximum one by 
\begin{eqnarray}
 \label{EQ19}
\displaystyle
  \begin{cases}
    F_{{\rm min}} &= \cfrac{G_{{\rm min}}-\frac{1}{2}\varepsilon (E_{{\rm glmin}})}{G_{{\rm max}}+\frac{1}{2}\varepsilon (E_{{\rm glmax}})}~, \\
  \rule[0mm]{0mm}{6mm}
    F_{{\rm max}} &= \cfrac{G_{{\rm min}}+\frac{1}{2}\varepsilon (E_{{\rm glmin}})}{G_{{\rm max}}-\frac{1}{2}\varepsilon (E_{{\rm glmax}})}~.
  \end{cases}
\end{eqnarray}
%Here, $\varepsilon(E)$ is the error calculated by standard error estimation, and 
Here, $\varepsilon(E)$ is the standard error of $G(E)$, and 
$E_{{\rm glmin}}$ and $E_{{\rm glmax}}$ are the energy values where 
$G_{{\rm min}}$ and $G_{{\rm max}}$ are obtained, respectively.
It is obvious that the value deteriorates as the size of system gets larger.
This means that it was difficult to estimate the DOS of large systems because of the large degrees of freedom.
%Here, $\varepsilon(E)$ is the error calculated by Eq.~(17), and 
%$E_{{\rm glmin}}$ and $E_{{\rm glmax}}$ are the energy values where 
%$G_{{\rm min}}$ and $G_{{\rm max}}$ are obtained, respectively.
%It is obvious that the value deteriorates as the size of system gets larger.
%This means that it was difficult to estimate the DOS of large systems because of the large degrees of freedom.
We needed more samples in order to obtain DOS for larger systems.
%*****************************
We could not find much differences among the three methods up to $N=16 \times 16$.
One bad data was founded in MUCAREM for the size of $N=32 \times 32$, 
and it made the error bar larger than the other methods.
If we consider the system larger than $N=32 \times 32$, REWL-MUCAREM gave the best results among the three methods.

In the present implementation of MUCAREM, we performed two multiple-histogram
reweighting (WHAM) operations: 
one after the REM simulation in the first half of the run and second after the MUCAREM simulation 
in the second half of the run.
Although a second WHAM operation converges quickly, the calculation cost of the 
first WHAM operation can become non-negligible in large systems.
The REWL-MUCAREM uses only the second WHAM and this WHAM converges even more quickly than in MUCAREM, because a good estimate of DOS is already prepared by the preceding REWL simulation.
Hence, REWL first and MUCAREM second is the order that we want to adopt in REWL-MUCAREM.
Note also that in REWL-MUCAREM, we do not need the piece-connecting process of DOS required in REWL, because WHAM automatically gives DOS in the entire energy range of interest.

%We can obtain good estimate of DOS under appropriate conditions and the DOS becomes more accurate by iterating the MUCAREM and WHAM reweighting operations.
%Although the most suitable conditions will depend on systems and methods, 
%the combination of REWL and MUCAREM can give accurate DOS with less calculation cost 
%(less number of replicas, less total number of MC sweeps).
%========================================
\section{Conclusions}
%\hspace{8mm}
%In this article, we investigated the two existing methods, MUCAREM and REWL, to 
%estimate the density of states in large systems.
In this article, we investigated an effective simulation protocol to 
estimate the density of states with highest accuracy.
We proposed REWL-MUCAREM that combines the advantages of REWL and MUCAREM, 
where REWL is performed first and MUCAREM is performed next.
This protocol was compared with existing two methods in a square-lattice 
Ising model, and 
the results showed that REWL-MUCAREM gave the most accurate density of states.

REWL-MUCAREM is effective with other systems and it can be extended to the 
MD simulation, 
because MUCA MD \cite{R2b,R24} and WL MD \cite{R25} have already been developed.
We have already calculated the residual entropy of Ice Ih \cite{R22} 
and we applied the protocol to helix-coil transitions of homo-polymers \cite{R23}.
The pre-views of these results are given in Appendix B.
REWL-MUCAREM MD simulations of protein folding are now under way.
We remark that there is an article which says that improvements can be 
transfered between MC and MD broad histogram methods~\cite{RESTMD_META}.
The protocol of REWL-MUCAREM is easy to implement and it can give more reliable results.\\
  
%A lot of advanced methods which based on the concept of MUCA and WL have been proposed.
%For example, STMC (Statistical Temperature Monte Carlo) or STMD (Statistical Temperature Molucular %Dynamics), which method have the concept of MUCA MD and WL MC, have been applied to spin grass %model and protein folding problem\cite{STMD,RESTMD,RESTMD_META}.
%In the terms of MD simulations of biopolymer, 
%Umbrella sampling\cite{UMBRELLA} and Metadynamics algorithm\cite{META}, which are interpretation %of MUCA and WL in reaction coordinates instead of potential energy, are preferred because reaction %coordinates are sometimes suitable in order to examine the changes of proteins.

%The protocol of REWL-MUCAREM can be easy to implement and it can give you more reliable results.\\
\noindent
{\bf Acknowledgements:}
%\hspace{8mm}
     
Some of the computations were performed on the supercomputers at the 
Supercomputer Center, Institute for Solid State Physics, 
University of Tokyo.\\
~~\\
~~\\
%\newpage
%\appendix
\centerline{{\bf Appendix A: Optimization of Conditions in MUCAREM}}
\setcounter{figure}{0}
\setcounter{table}{0}
\setcounter{equation}{0}
\setcounter{section}{1}
\renewcommand{\thefigure}{\Alph{section}\arabic{figure}}
\renewcommand{\thetable}{\Alph{section}-\Roman{table}}
\renewcommand{\theequation}{\Alph{section}\arabic{equation}}
%\section{Specific heat}
%*****************************
%\begin{figure}[tbhp]
%  \centering
 %\includegraphics[width=8.0cm]{SPHEAT.png}
%\includegraphics[width=8.0cm]{SPHEAT.eps}
% \includegraphics[width=12.0cm]{SPHEAT.eps}
%  \caption{The specific heat. (a) is the exact solutions which were calculated by exact DOS\cite{R17,R18}.
%(b), (c), and (d) were obtained by simulations with $L=8,$ 16, 32, 64, and 128 
%by MUCAREM, REWL, and REWL-MUCAREM, respectively.}
%\end{figure}
%*****************************

%\section{Selection}
%========================================
%\begin{figure}
%\begin{centering}
% \includegraphics[width=12.0cm,height=10.0cm]{DOS.eps}
%\end{centering}
%\end{figure}
%========================================
%\begin{figure}
%  \centering
% \includegraphics[width=12.0cm]{FLATNESS.eps}
%According to B.A. Berg, $F$ is enough to estimate the expectation values if it takes 0.5.}
%\end{figure}
%========================================

%\section{Optimization of Conditions in MUCAREM}
%========================================
%\begin{table*}[b]
\begin{table*}[t]
\caption{\label{tab:table2} Conditions of MUCAREM simulations.
%\textbackslash
}
%\begin{ruledtabular}
\begin{tabular}{cccccccc}
 \hline 
 \hline 
      Methods
   & \, Number of spins \,
   & \, MUCAREM \,
   & \multicolumn{2}{c}{\, Number of replicas \, }
   & \multicolumn{2}{c}{\, Number of MC sweeps \,}
   & Total MC sweeps       \rule[0mm]{0mm}{4mm} \\
     
   & $N$
   & iterations
   & \multicolumn{2}{c}{}
   & \multicolumn{2}{c}{per replica}
   & \\
 \hline 
   &
   &
   & \, REM \rule[0mm]{0mm}{4mm}
   & MUCAREM
   & REM
   & MUCAREM
   &      \\
 \hline
MUCAREM1   &64&1&4&4&100\,000&100\,000&~$200\,000 \times 4$ \rule[0mm]{0mm}{4mm}\\
                   &256& &8&8&100\,000&100\,000&$200\,000 \times 8$ \\
                   &1024& &16&16&100\,000&100\,000 &$200\,000 \times 16$ \\
                   &4096& &32&32&250\,000&250\,000&$500\,000 \times 32$ \\
                   &16384& &64&64&1\,500\,000&1\,500\,000&$3\,000\,000 \times 64$ \\
MUCAREM2 &64&1&8&4&10\,000&180\,000&$200\,000 \times 4$ \\
                   &256& &16&8&10\,000&180\,000&$200\,000 \times 8$ \\
                   &1024& &32&16&10\,000&180\,000&$200\,000 \times 16$ \\
                   &4096& &64&32&25\,000&450\,000&$500\,000 \times 32$ \\
                   &16384& &128&64&150\,000&2\,700\,000&$3\,000\,000 \times 64$ \\
MUCAREM3 &64&2&8&4&10\,000&90\,000&$200\,000 \times 4$ \\
                   &256& &16&8&10\,000&90\,000&$200\,000 \times 8$ \\
                   &1024& &32&16&10\,000&90\,000&$200\,000 \times 16$ \\
                   &4096& &64&32&25\,000&225\,000&$500\,000 \times 32$\\
                   &16384& &128&64&150\,000&1\,350\,000&$3\,000\,000 \times 64$ \\
 \hline 
 \hline 
\end{tabular}
%\end{ruledtabular}
\end{table*}
%========================================
We would like to discuss the optimization conditions in MUCAREM in order to obtain more accurate DOS in REWL-MUCAREM.
We performed two more MUCAREM simulations with different conditions. 
Table A-I lists the conditions of the additional simulations (MUCAREM$2$ and MUCAREM$3$) together with the first MUCAREM in Table I
(which is now referred to as MUCAREM$1$).
The major differences of the additional MUCAREM simulations from the previous MUCAREM simulation lies in the following: number of MC sweeps for REM and MUCAREM, number of replicas used for REM,  and number of iterations of MUCAREM.
In the additional MUCAREM simulations, the $10~\%$ of the total MC sweeps was run with REM 
and the remaining $90~\%$ of the simulation was MUCAREM with the DOS obtained from the preceding REM simulation.
In REM simulations in MUCAREM$2$ and MUCAREM$3$, replica exchange was proposed every $10$ MC sweeps.
On the other hand, in MUCAREM in MUCAREM$2$ and MUCAREM$3$, replica exchange was proposed every $100$ MC sweeps.

It is often said that the DOS obtained by MUCAREM simulations becomes better by iterating MUCAREM and WHAM reweighting\cite{R10}.
We iterated MUCAREM simulations once for MUCAREM$3$. 
%Twenty-five independent runs with different initial random number seeds were performed.
%(i.e., $n=25$, and we obtained estimated DOS $25$ times).
It should be mentioned that because we obtained clearly wrong DOS, we performed one extra run for each system $N=32 \times 32$ and $N=128 \times 128$ in MUCAREM$2$, and simply discarded these apparently bad runs.
(We did not find a bad run in REWL and REWL-MUCAREM simulations.)
We think the problem came from the difficulty of uniformly sampling over a wide energy range in REM.
The rough DOS obtained from the first REM was not good and the inaccuracy had a bad influence to the sampling of the following MUCAREM.
Giving apparently wrong results suggests that MUCAREM simulations are unstable comparing to REWL and REWL-MUCAREM simulations, which implies that the total number of MC sweeps and/or the number of runs $n$ should be longer for MUCAREM simulations to conclude with confidence.
%*****************************
\begin{figure}[thbp]
%  \begin{center}
   \includegraphics[width=120mm]{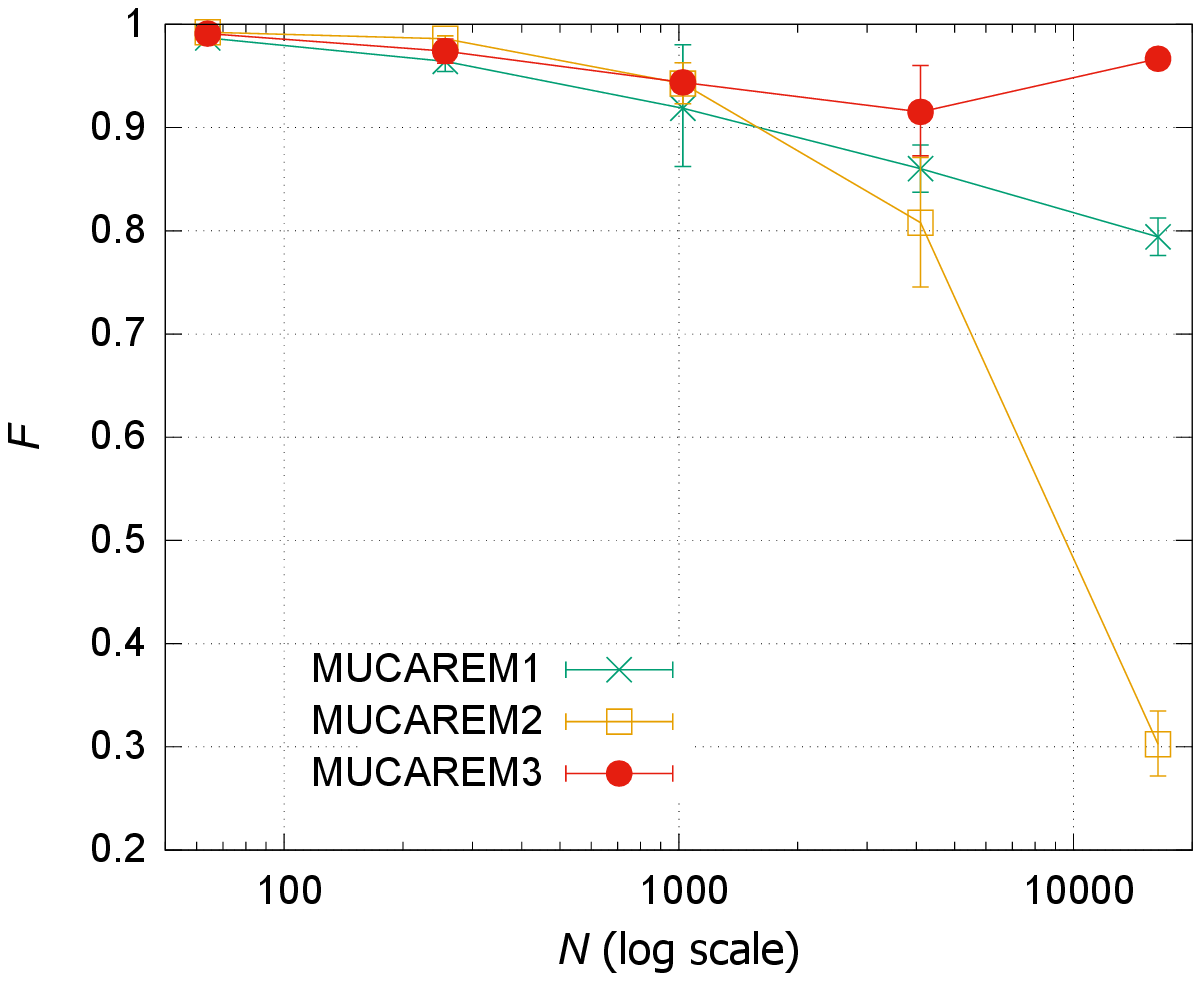}
%  \end{center}
  \caption{Global flatness $F$ defined in Eq.~(\ref{EQ18}) as a function of the total number of spins, $N$.　The error bars of MUCAREM3 except for $N=64 \times 64$ were sufficiently small compared to the symbols.}
% \end{minipage}
\end{figure}
%*****************************
%Fig.~$5$ shows the results of mean local flatness $G$ in Eq.~$(16)$ for the system $N=32 \times 32$.
%It is obvious that MUCAREM$2$ and MUCAREM$3$ gave the DOS with smaller error bars than MUCAREM$1$.
%This means that we can estimate more accurate DOS if the conditions are selected carefully.
%It is worth saying that the width of the error bars of MUCAREM$2$ and MUCAREM$3$ were almost the same as in REWL-MUCAREM in Fig.~$3$.

Fig.~A1 shows the global flatness $F$ in Eq.~$(\ref{EQ18})$.
Although there are little differences in $F$ among the three MUCAREM 
simulations up to the system $N=64 \times 64$, we found a large difference 
in the system $N=128 \times 128$.
MUCAREM$3$ could obtain a good estimate of DOS compared with MUCAREM$1$ and MUCAREM$2$.
It implies that the DOS became better by iterating MUCAREM simulations rather than by using more sampling obtained from a single, long run of MUCAREM.
The error bar for the system $N=64 \times 64$ is large if it was compared to other size of systems in MUCAREM$3$.
We found a bad result from one run out of the $25$ runs, which made the error bar larger.
These again suggest that MUCAREM simulations are unstable compared to REWL and REWL-MUCAREM simulations.
Note that the estimation of DOS under the conditions of MUCAREM$3$ for $N=128 \times 128$ is even better than that of REWL-MUCAREM in Fig.~$3$, although MUCAREM seems to be unstable.
We expect that REWL-MUCAREM could remove the instability of MUCAREM and it would give better DOS if its MUCAREM simulation in REWL-MUCAREM was also iterated.

We can obtain good estimate of DOS under appropriate conditions and the DOS becomes more accurate by iterating the MUCAREM and WHAM reweighting operations.
Although the most suitable conditions will depend on systems and methods, 
the combination of REWL and MUCAREM can give accurate DOS without worrying about bad data.

%*******************************
%*******************************
%~~\\
%~~\\
\newpage
\centerline{{\bf Appendix B: Applications of REWL-MUCAREM}}
~~\\
\setcounter{figure}{0}
\setcounter{table}{0}
\setcounter{equation}{0}
\setcounter{section}{2}
%*****************************
We applied the REWL-MUCAREM MC protocol to two characteristic systems, 
Ice Ih and biopolymer. 
In this Appendix, we show some preliminary results.

The first example is the estimation of the residual entropy of Ice Ih 
by REWL-MUCAREM \cite{R22}.
According to Pauling's theory, the residual entropy is obtained from 
the degrees of freedom 
of orientations of water molecules which are observed in the 
groundstate \cite{ICE_PAU}.
Two simple Potts-like models, which are referred to as the 2-site model 
and 6-state model, with nearest-neighbor interactions 
on three-dimensional hexagonal lattice were introduced and the 
residual entropy was estimated by a MUCA simulation \cite{R16}.
%(The details of these models were mentioned in~\cite{R16}.)
%These models are associated with three-dimensional Potts-like model.
We applied our protocol to the 2-site model for obtaining the residual 
entropy with high accuracy.
%It is preferred to obtain the candidates of the residual entropy in 
%large systems because 
The final estimation of the residual entropy is estimated by extrapolation,
taking the thermodynamic limit.

The estimations of the degrees of freedom of orientations of one water 
molecule $W_{1}(1/N)$ are shown in Fig.~B1(a).
Here, $N$ stands for the total number of water molecules of the system.
The relationship between the degrees of freedom $W_{1}(1/N)$ and the 
residual entropy $S_{1}(1/N)$ is given by
\begin{eqnarray}
 \label{BEQ1}
\displaystyle
 S_{1} \left(\cfrac{1}{N}\right) = k_{{\rm B}} \log W_{1}\left(\cfrac{1}{N}\right)~.
\end{eqnarray}
In Fig.~B1(a), three data points ($N=1600, 2880, 4704$) are plotted.
It was not possible to obtain the value for
$N=4704$ by the previous MUCA simulations \cite{R16,ICE2012},
and we needed the REWL-MUCAREM to obtain this new value.
A fit (the green curve in Fig~B1(a)) for the data to the form
\begin{eqnarray}
 \label{BEQ2}
\displaystyle
 W_{1}\left(\cfrac{1}{N}\right) = W_{0} + a_{1}~\left(\cfrac{1}{N}\right)^{\theta}~
\end{eqnarray}
is shown. 
Here, $W_{0}$, $a_{1}$ and $\theta$ are fitting parameters 
and $W_{0}$ converts to 
the final estimation of the residual entropy $S_{0}$ in the
thermodynamical limit ($N \rightarrow \infty$).
%Note that we used additional $5$ data of $W_{1}(1/N)$ calculated 
%in smaller systems ($N=128,288,360,576,896$) for the fit.

Our final estimation of $W_{0}$ is \cite{R22}
\begin{eqnarray}
 \label{BEQ3}
\displaystyle
W_{0} &= 1.507480 \pm 0.000048~,
\end{eqnarray}
and the final residual entropy $S_{0}$ is
\begin{eqnarray}
 \label{BEQ4}
\displaystyle
S_{0} = 0.815627 \pm 0.000039~~~\left[{\rm cal/deg~mole}\right].
\end{eqnarray}
These results agree well with the other results in~\cite{ICE_J,ICE_V}, 
which were estimated by other simulation methods.
%(e.g. Thermodynamic Integration).
%We would like to discuss the results and the analysis in~\cite{R22}.

%*****************************
\begin{figure}[tbh]
\begin{tabular}{cc}
\centering\includegraphics*[width=75mm,height=75mm,keepaspectratio,clip]{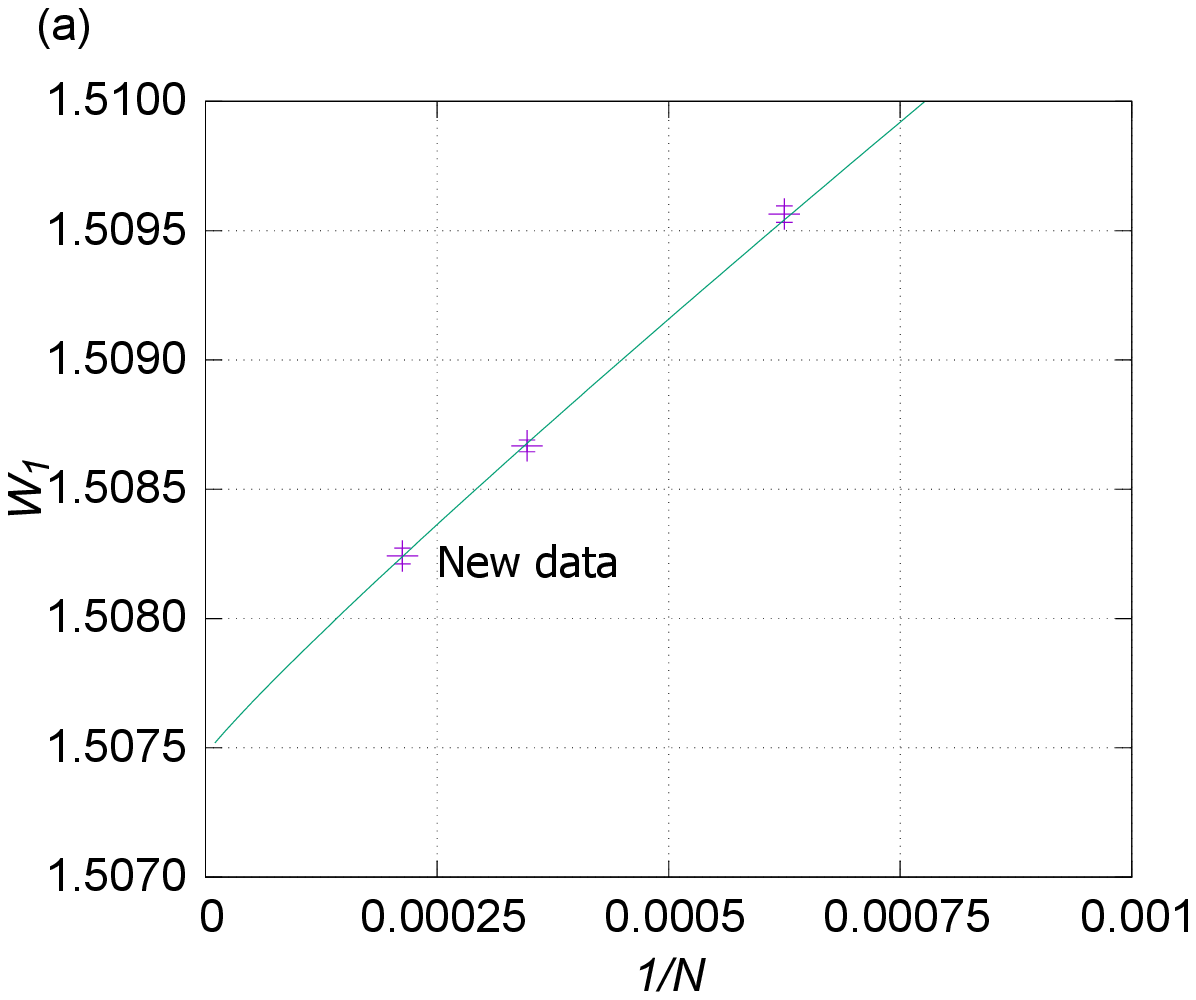}
%&&
\centering\includegraphics*[width=75mm,height=75mm,keepaspectratio,clip]{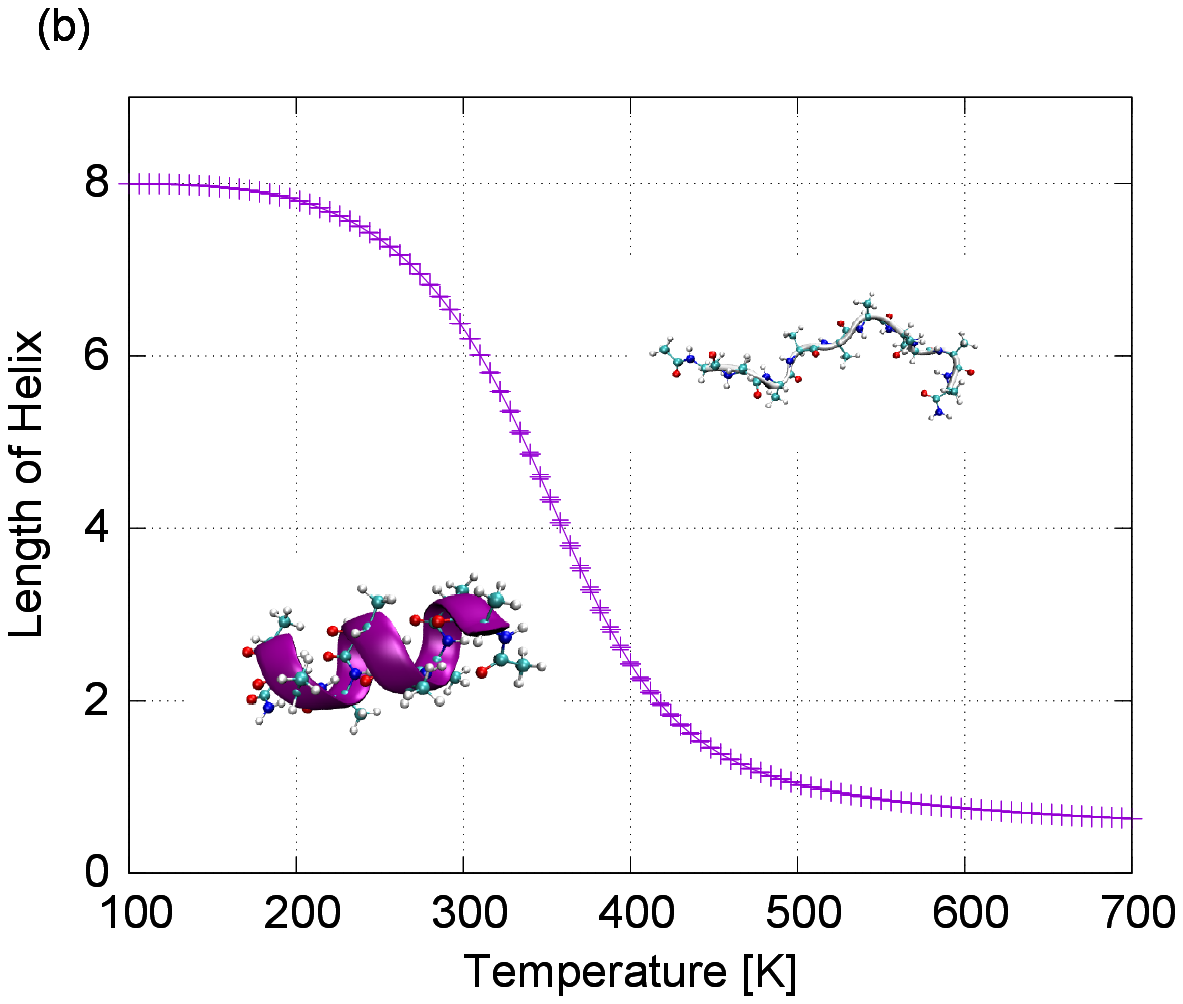}
\end{tabular}
 \caption{(a)~The number of degrees of orientation per water molecule 
$W_{1}$ as a function of the number of total water molecules.
The purple data points correspond to $N=1600, 2880, 4704$. The green fit 
curve was estimated by Eq.~(\ref{BEQ2}). (b)~The temperature dependence 
of average helix length of deca-alanine. Deca-alanine takes a coil structure 
at high temperatures (see the right structure) and takes an $\alpha$-helix 
structure at low temperatures (see the left structure).}
\end{figure}
%*****************************
Another example is a folding simulation of a simple biopolymer \cite{R23}.
%It is well known that the three-dimensional structures of proteins are 
%closely related to their functions in vivo. 
%According to the Anfinsen's dogma, the structures are determined by 
%their amino-acid sequences. 
%One of the main questions of protein folding problems is how to predict 
%the three-dimensional structures 
%from its amino-acid sequence information.
%However, there is a problem for conventional
% MC or MD simulations of large and complex proteins.
%Powerful simulation methods (e.g. generalized-ensemble algorithms) 
%are essential for folding simulation 
%because of the complexity of energy landscape of proteins where 
%conventional MC or MD simulations 
%are easy to get trapped in the local state.
In order to examine the effectiveness of our protocol for protein 
folding simulation, 
we studied the helix-coil transition of a deca-alanine 
(which is a helix former) with AMBER99/GBSA force field.
REWL-MUCAREM MC protocol was employed and the dihedral angles between 
residues were updated 
by the Metropolis criterion during simulations.

The temperature dependence of the average helix length of deca-alanine 
is shown in Fig.~B1(b).
%The horizontal axis shows the temperature and the vertical axis shows 
%the length of helix.
Because the structures of the terminal residues 
%at N terminus and C terminus are discarded, 
are frayed, the maximum helix length is 8.
The deca-alanine is in a coil state above the transition temperature 
of $T_{c}~\approx~350$ K and in a helix state below $T_{c}$.
%The results means that REWL-MUCAREM was applied to protein 
%folding simulation successfully.
Folding simulations of larger and more complex proteins are under way
\cite{R23}.
%We intend to provide the results and analysis in~\cite{R23}.
%\newpage
%========================================

%========================================

%
\end{document}